\documentclass[%
 reprint,
 twocolumn,
 showpacs,
 showkeys,
 preprintnumbers,
 amsmath,amssymb,
 aps,
  pra,
  longbibliography,
 ]{revtex4-1}

\usepackage{natbib}

\usepackage{tikz}
\usepackage[breaklinks=true,colorlinks=true,anchorcolor=blue,citecolor=blue,filecolor=blue,menucolor=blue,pagecolor=blue,urlcolor=blue,linkcolor=blue]{hyperref}
\usepackage{graphicx}
\usepackage{url}

\usepackage{xcolor}

\usepackage{eurosym}

\usepackage[none]{hyphenat}

\begin{document}
\sloppy

\title{Quantum Hocus Pocus}


\author{Karl Svozil}
\affiliation{Institute for Theoretical Physics, Vienna
    University of Technology, Wiedner Hauptstra\ss e 8-10/136, A-1040
    Vienna, Austria}
\affiliation{Department of Computer Science, University of Auckland, Private Bag 92019,  Auckland 1142, New Zealand}
\email{svozil@tuwien.ac.at} \homepage[]{http://tph.tuwien.ac.at/~svozil}


\begin{abstract}
The claims made in a manifesto resulting in the European quantum technologies flagship initiative in quantum technology and similar enterprises are taken as starting point to critically review some potential quantum resources, such as coherent superposition and entanglement, and their potential usefulness for parallelism and communication. Claims of absolute, irreducible (non-epistemic) randomness are argued to be metaphysical. Cryptanalytic man-in-the-middle attacks on quantum cryptography are well known to be feasible, but hardly mentioned. If all of this is taken into account, a more sober perspective on quantum capacities emerges, but one that may be ethically more justified than the ``hype and magic'' that drives many current initiatives.
\end{abstract}

\maketitle

\section{Introduction}

The European  Quantum Manifesto~\citep{2016-05-qmanifesto}
contributed to the launch of a
\euro{}1 billion quantum technologies flagship initiative in quantum technology~\citep{2016-05-EU-qi}.
Thereby, quanta which {\em ``can be in different states at the same time (`superposition')
and can be deeply connected without direct physical interaction (`entanglement')''}
are expected to create a {\em ``second quantum revolution''} by
{\em ``taking quantum theory to its technological consequences''}~\citep{2016-05-EU-qr}.

This is in line with assurances by many proponents that ``quantum mechanics is magic'';
and indeed so irreducible incomprehensible
by rational human thought that anybody asking~\citep[p.~129]{feynman-law} {\em ``But how can it be like that?'' will be dragged
``~`down the drain', into a blind alley from which nobody has yet escaped.''}
From that perspective, it appears prudent to harvest these alleged capacities beyond classical algorithmics
for technology and the economy at large;
in particular if the experts proclaim such a program to be feasible.

I would like to state up-front that I am not criticising this new initiative on grounds that money will be wasted.
On the contrary, it will be money wisely spent,
and many new and interesting research and technological developments will spin off.

However, in what follows I would like to point out that, at least in the way it is marketed,
the quantum technologies flagship initiative in quantum technology is deceptive,
if not dangerously misleading.

It is deceptive because, while many of the   Quantum Manifesto's short- and medium-term goals are reachable or have already been achieved,
some of these goals strongly depend on the assumptions made.

It is dangerous because it pretends to deliver -- for instance, with respect to quantum random number generators
and quantum cryptography -- what is provable impossible.

Moreover it is highly unlikely that some of the long-term goals are achievable even in principle.

\section{Quantum computation}

Let us first review quantum computation; in particular the Quantum Manifesto's
long-term goal to {\em ``Build a universal quantum computer able to demonstrate the resolution of a problem that, with
current techniques on a supercomputer, would take longer than the age of the universe''}~\citep[p.~18]{2016-05-qmanifesto}.
I do not know what the authors had in mind by formulating this bold claim, but
when it comes to quantum computation as compared to classical universal computation there are at least two issues
which have to be kept in mind: one is algorithmic in nature, and one is hardware related.

\subsection{Quantum algorithm}

As anecdotal as this may sound, one of the greatest former talents in quantum computation,
and co-author of an authoritative volume
on the subject (which, after publication in 2000, made it to the 10th anniversary edition in 2010)
gave up his tenured academic position {\em ``in order to work as an advocate for open science''}~\citep{nielsen-blog}.

Such brain drain is surprising, given the hype.
Alas it might not be too negative to state that,
besides a growing zoo of quantum algorithms~\citep{jordan-zoo}
(and notwithstanding some progress in communication complexity~\citep{Raz:1999:ESQ:301250.301343,Montanaro:2011:NES:2230916.2230919,Gavinsky:2016:ESV:2897518.2897545},
given unlimited computational power)
quantum algorithms have not much advanced since the proposal of Grover's algorithm, for a period of twenty years now.
So, to call this field of research ``progressive'' might be overly optimistic.

Moreover, while quantum factoring is often mentioned as ``killer app'' for quantum computation,
classical prime factorization is neither in the class of NP-complete problems,
nor can it be excluded that classical algorithms solve this task in polynomial time; just like Shor's probabilistic quantum algorithm.


The key issue, in my opinion, is a lack of knowledge of just what the quantum assets and capacities,
capable of potentially trespassing classical computational means, really are.

\subsubsection{Parallel processing by superpositions}

Many researchers would be inclined to postulate quantum superpositions -- the capacity to simultaneously co-represent classically distinct, even mutually contradictory,
states -- and the resulting sort of parallelism,
as one of the main quantum-over-classical advantages.

Unfortunately, all of our attempts to comprehend a widely cited paper on quantum complexity theory~\citep{be-va} failed.
In particular, their hint in Sect.~3.3, that superposition (and thus parallelism) requires a huge (exponential)
computational capacity (one that could potentially be harvested) of the physical universe,
is immediately questioned by mentioning restrictions due to the fact that the quantum evolution
is essentially a permutation of the quantum state.

A recent review~\citep{Montanaro:2016:npjqi.2015.23} also attempts to locate
quantum capacities by emphasizing coherent superpositions (and thus parallelism).
It is mentioned that
a cynical reader might point out that, based on a result by
Shi~\citep{Shi:2003:BTC:2011508.2011515} any quantum algorithm
whatsoever can be expressed as the use of just two components:
(i) gates producing coherent superpositions of a classical bit  (such as the Hadamard gate
or quantum Fourier transforms), interspersed with (ii) classical processing.

Alas, all the parallel ``results'' of a quantum computation encoded in a coherent superposition
are not directly accessible: due to quantum complementarity and the no-cloning theorem
there is no way to access and measure complementary aspects of an arbitrary pure state comprehensively.
In terms of the many-worlds interpretation,
every one of the parallel results resides in one of those parallel worlds simultaneously;
but any particular observer has direct access to only one such universe.

Indeed, relative to ``reasonable'' assumptions, observables which are not identical to pure states (and their negation),
cannot consistently (co-)exist with the latter~\citep{pitowsky:218,2015-AnalyticKS}.
From this point of view, ``coherent superpositions'' just correspond to improper, misleading representations of non-existing aspects of physical reality.
They are delusive because they confuse ontology with epistemology~\citep{jaynes-89,jaynes-90}
by suggesting the physical co-existence of counterfactuals; in particular, classically inconsistent cases,
in an exploitable classical manner.
However, upon closer inspection, this alleged capacity might just be a consequence of a misconception
yielding an operational ill-representation of the quantum state~\citep{svozil-2013-omelette}.

``Forcing'' a ``measurement'' of such states in a coherent superposition of ``observables''
results in a context translation~\citep{svozil-2003-garda}. This may introduce stochasticity
due to the many
(for all practical purposes~\citep{bell-a}) uncontrollable degrees of freedom of the measurement device~\citep{engrt-sg-I}.

Nevertheless, with all these provisos, a potential quantum advantage resides in the possibility to encode
certain suitable relational functional properties representable by
(equi-)partitions of the image of the function~\citep{DonSvo01,svozil-2002-statepart-prl}
into suitable orthogonal projections~\citep{svozil-2016-vector}.
Unfortunately this is not ubiquitous,
as for certain tasks such as parity effective speedups are impossible~\citep{Farhi-98}.

\subsubsection{Multipartite communication by entanglement}

Quantum mechanics denies the separate existence and apartness of certain entities (such as quanta of light)
``tightly bundled together'' by entanglement.
Indeed, the entire state of multiple quanta
can be expressed completely, uniquely and solely in terms
of correlations (joint probability distributions)~\citep{Bergia1980,mermin:753},
or, by another term, relational properties~\citep{zeil-99},
among observables belonging to the subsystems;
irrespective of their relativistic spatio-temporal locations~\citep{Seevinck:2010eb}.
Consequently~\citep{Bennett-IBM-03.05.2016}
one has ``a complete knowledge
of the whole without knowing the state of any one part. That a thing can be in a definite
state, even though its parts were not.''

In more technical terms, this can be interpreted as just another consequence of quantum coherence
--
only that the co-represented classical cases refer to product states of multiple quanta,
thereby effectively allowing two or more different quanta to
be coherently connected over a large distance.
Note also that if the two parties share
correlated pairs of quanta, then (by quantum teleportation)
the quantum communication and selection between those parties can be done by classical information.

While it may be too early for a definite answer, many (exponential) quantum speedups~\citep{Raz:1999:ESQ:301250.301343,Montanaro:2011:NES:2230916.2230919,Gavinsky:2016:ESV:2897518.2897545}
might again, just as in the functional case, be due to the possibility to encode communication tasks into suitable orthogonal subspaces.
Observe that every binary function
$g: \textsf{\textbf{x}} \times \textsf{\textbf{x}} \rightarrow \textsf{\textbf{z}}$
can be converted into an equivalent unary function
$f: \textsf{\textbf{x}} \rightarrow \textsf{\textbf{z}}^\textsf{\textbf{x}}$,
such that $g(\textsf{\textbf{x}}_1,\textsf{\textbf{x}}_2) = [f (\textsf{\textbf{x}}_2)](\textsf{\textbf{x}}_1) \in \textsf{\textbf{z}}$.
One may think of
$\textsf{\textbf{x}}_2$ as some ``index'' running over unary functions $f$.
If this ``index'' can be efficiently communicated, $f$ and its equivalent representation $g$ can be evaluated.

\subsection{Quantum hardware}

In the last thirty years single-quantum experiments,
such as single quanta in a double slit, and all kinds of other interference and state (re-)construction experiments
sharpened and enlightened our understanding of the quanta.
One of the main features of the (unitary quantum) evolution is that it is a permutation of the state;
therefore, at least in principle, information can neither be created (or copied), nor lost.
Thus designs of quantum computers have to answer the question of how to get rid of auxiliary qubits (they cannot).

Another formidable question is to maintain coherence over sufficient amounts of computation space and time,
thereby keeping the system isolated; that is, by avoiding entanglement with the environment.
It may well be that maintenance of coherence scales exponentially with both computation space and time,
thereby rendering quantum computation non-scalable.

It should be kept in mind that while it may, in very special cases,
be possible to obtain quantum coherence for more than a thousand qubits,
those systems are non-universal, and specifically tailored for very particular tasks.
And, of course, every system is a perfect simulation of itself; so as every system is quantized,
it is also a perfect simulation of a multipartite quantum state --
indeed, this could involve zillions of quanta.

\section{Hypercomputational capacities through irreducible quantum randomness}
\label{2016-quantum-hokus-pokus-s-qr}

Since ``true'' sources of randomness are often required in quantum information theory such as in quantum
cryptography, quantum random number generators will be shortly discussed next.
While Born~\cite{born-26-1} and others have expressed their personal inclinations about randomness in nature,
and explicitly stated their very subjective choices as such,
this supposition has been canonized and postulated as an axiom.
It is corroborated by our obvious inability to come up with theoretical predictions of certain quantum outcomes.

In practice, quantum random number generators
are tested and certified by performing a battery of statistical criteria, such as diehard tests, on finite sequences of data.
This is far from the claims of absolute, irreducible certification promised to customers.

Unfortunately, by merely studying the raw data without additional assumptions (such as the quantum axiom mentioned),
and even if the supposedly random data sequences could be provided at arbitrary length,
due to the recursive undecidability of the rule inference problem, and other theorems of recursion theory,
claims of absolute randomness are provable unprovable; and therefore are metaphysical,
that is, beyond the reach of science.
In other words, science can neither assert nor disprove quantum randomness, and never will be able to do so:
this method is (provably) blocked by limits due to consistency, and consequently the avoidance of paradoxical self-reference~\citep{Yanofsky-BSL:9051621}.

Therefore, any claims that quantum random number generators are certified
by the very laws of nature to behave indeterministically
are incorrect.
Certification resides in, and is relative to,
the validity of canonical quantum theory,
which in turn resides in our beliefs in it.

Another way of thinking about quantum randomness is in terms of the -- supposedly (that is, relative to the axioms)
``indeterministic'' generation process.
Particular single outcomes are thought of as occurring without deterministic cause; quasi {\it ex nihilo}.
In theological terms such outcomes are by  {\it creatio continua}.
Thereby the ``measurement of the outcome'' is postulated to come about in a quantum formalism based on an evolution that one-to-one permutes the state
(which consequently has a unique history);
an ambivalence~\citep[p.~454]{everett} which is protected through orthodoxy.

Very often, it is also not explicitly disclosed how exactly such random sequences are generated
and where the randomness resides
-- would, for instance, a source of photons impinging on two detectors qualify as beam splitter?
-- not to mention the fact that lossless beam splitters are represented by one-to-one unitary transformations, that is, merely permuting the state;
let alone the method of normalization of the unbiased raw signals.

\section{Quantum cryptography}

With regards to cryptography, the Quantum Manifesto mentions two goals; one medium-term:
{\em ``Enable secure communication between distant cities via quantum networks, which enhance information
security and make eavesdropping impossible''}~\citep[p.~17]{2016-05-qmanifesto};
as well as one long-term: {\em ``Create a secure and fast quantum internet connecting the major cities in Europe using quantum
repeaters running quantum communication protocols.''}~\citep[p.~18]{2016-05-qmanifesto}

Contrary to publicized claims, quantum cryptography is insecure and can be successfully cryptanalyzed through man-in-the-middle attacks,
that is, by compromising both quantum and (public) classical communication lines
(cf. Ref.~\citep{svozil-2005-ln1e} for a ``demonstration'' using Viennese chocolate balls).
This is a well known fact which is already explicitly mentioned in the original paper by
Bennett and Brassard~\citep{benn-84} as follows:
{\em ``The need for the public (non-quantum) channel
in this scheme to be immune to active eavesdropping
can be relaxed if Alice and Bob have agreed beforehand
on a small secret key, which they use to
create Wegman-Carter authentication tags [WC] for
their messages over the public channel.''}

Alas, the consequences of the cryptanalytic capacities that can be deployed through non-immune classical
channels are more devastating than they first may appear.
Because quantum cryptography is often seen as a remedy for non-immune, and thus compromised, public classical channels.
However, in order to prove ``unconditional security'' of quantum cryptography,
it has to be assumed that the public classical channel is immune (thus the necessity of classical authentication).
As a consequence, one is relegated to ``growing'' an initial key at best; but key growing
might be perceived as merely a gradual improvement over classical methods, since the identities
of the communicating parties still need to be checked by classical authentication.

In short, if a classical channel is not compromised, no quantum cryptography is required.
And since quantum cryptographic protocols such as the one mentioned earlier presuppose an immune classical channel,
they can be compromised if the classical channel is compromised.

This simple fact is often ``taken for granted'' and not mentioned in
proofs of ``unconditional security of quantum cryptography;'' even in
authoritative reviews of the subject.
As a consequence, those proofs are correct relative to the absence of
tampering with the classical channel.

One of the problems with claims of absolute security (certified by the quantum nature) is that,
as in other domains,
while ignorance favours the proponents of a technology,
the real costs as well as the disadvantages have to be borne by others having their skin in the game.
Ernst Specker called such particular instances of non-disclosure ``Jesuit lies'' because
neglect to mention allegedly obvious but important and decisive unfavourable facts is different
from stating false propositions; Jesuits have faced a not dissimilar problem (and solution)
under torture or danger.

By the same rhetoric fission reactors are `unconditional secure,'
provided earthquakes and tsunamis are absent; as well as reckless misconduct
and other problems that would make them insecure.

\section{Whatever it takes}

As I have emphasized at the beginning, I have no intention to criticize the European flagship initiative in quantum technology
on grounds that liquidity is poured into certain quantum laboratories and industries.
What I criticize is the hubris in marketing it.
Of course one might say that, at the end of the day, nobody will remember the claims that initiated the funding,
all of its proponents and political supporters and enablers will be gone,
and many valuable findings and technologies will spin off from it anyhow.
After all, one has to exaggerate in order to motivate and account for resource allocation in societies like ours.

I believe that science will fare better if it goes for the (sometimes ``awful'' or complicated)
truth in the long run, and not for marketable promises.
It should be made clear to the public at large what are the stakes and realistic prospects,
and what are the risks of funding;
rather than trumpeting out vague claims which deceive and serve expectations rather than inform.
In the public interest, as well as for scientific progress,
funding agencies and scientific organizations need to
allocate more space, time and resources to ``negative'' contributions~\citep{mueck-nn-13} which are critical about feasibility and status;
in particular, when it comes to conference contributions and publications.

Let me finally express one opinion about a research area that I find positively necessary to finance:
nuclear fusion research. In view of the energy crisis that will affect and deeply transform our societies in the not-so-far future,
we need to make sure that we have sufficient electric energy deployable, which could eventually substitute
the depleting oil reserves.
I believe it is not overstated that, despite the tremendous challenges and
obstacles in physics and material science of this prospective technology,
thermonuclear fusion reactors could provide us with the energy our societies need; accompanied
with sustainably bearable side effects. At the moment the two formidable problems  --
creating an environment for fusion, as well as being able to thermalize the energy released during fusion
in a sustainable manner
--
might require a commitment that goes far beyond the \euro{}1 billion input into the quantum technologies flagship initiative in quantum technology
discussed here.
But, as this might become a necessity rather than a convenience in the medium-term future,
we should spend ``whatever it takes'' to accomplish this energy goal;
regardless of the price of energy today, thereby transforming the petrochemical industry,
as well as our societies at large,
into entities that could survive and prosper during and after the upcoming energy crisis.

\acknowledgments{
This work was supported in part by the European Union, Research Executive Agency (REA),
Marie Curie FP7-PEOPLE-2010-IRSES-269151-RANPHYS grant.

Responsibility for the information and views expressed in this article lies entirely with the author.
The content therein does not reflect the official opinion of the Vienna University of Technology or the University of Auckland.

The author declares no conflict of interest and, in particular, no involvement in nuclear fusion research.

}


%

\end{document}